# KINEMATICAL AND DYNAMICAL APPROACHES TO GRAVITATIONAL INSTABILITY

EDMUND BERTSCHINGER
*Institute for Advanced Study, Princeton, NJ 08540 USA*
*Department of Physics, MIT 6-207, Cambridge, MA 02139 USA*

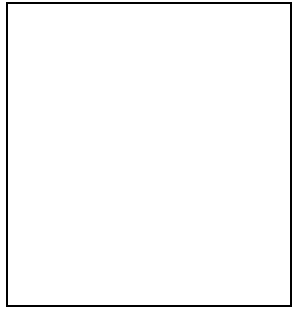

**Abstract**

This paper reviews the essential physics of gravitational instability in a Robertson-Walker background spacetime. Three approaches are presented in a pedagogical manner, based on (1) the Eulerian fluid equations, (2) the Lagrangian description of trajectories, and (3) the Lagrangian fluid equations. Linear and nonlinear limits are discussed for each case. Shear and tides are shown to play a key role in nonlinear gravitational instability.

The Lagrangian fluid approach is used to show that several widely held beliefs about gravitational instability are false. The following collapse theorem is proven: for a given initial density fluctuation and growth rate, the spherical tophat perturbation collapses more slowly than any other configuration. We also show that density maxima are not the first points to collapse and that underdense regions may collapse if their initial shear is sufficiently high. The Lagrangian fluid approach leads to an almost closed set of local evolution equations for individual mass elements. The magnetic part of the Weyl tensor, which may be present even in the nonrelativistic (Newtonian) limit, may prevent a purely local description. However, neglecting the magnetic part of the Weyl tensor, we obtain predictions for high-redshift collapse that are in good agreement with a high-resolution cold dark matter N-body simulation.

## 1 Introduction

Gravitational instability theory provides useful relations between the large-scale mass (or galaxy) density field $\rho(\vec{x})$ and velocity field $\vec{v}(\vec{x})$. These relations allow us, in principle, to achieve the following goals:

1. Reconstruct three-dimensional fields from distances and redshifts ($H_0 r$, $cz$) ([3])

2. Recover initial fluctuation fields ([26], [15])

3. Test theories of initial fields ([18], [11], [30], [28])

4. Test gravitational instability paradigm ([10])

5. Determine cosmological parameters ([19], [10])

This program constitutes a growing subfield of cosmological research.

Achieving these goals requires use of theoretical tools of gravitational instability. This paper summarizes some of the methods and recent results, with particular emphasis on the Lagrangian fluid method.

## 2 Eulerian fluid equations

The best-known dynamical description of mass is given by the Eulerian fluid equations. We express them here using the comoving position $\vec{x}$ and conformal time $\tau$, which are related to the proper position $\vec{r}$ and time $t$ measured in a locally inertial comoving frame by

$$\vec{x} = \vec{r}/a(t) , \quad d\tau = dt/a(t) , \tag{1}$$

where $a(t)$ is the cosmic expansion scale factor. Because $a > 0$, $\tau$ is a monotonic function of $t$, so that we may write $a = a(\tau)$. (E.g., for an Einstein-de Sitter universe, $a \propto t^{2/3}$ implies $\tau \propto t^{1/3}$ and $a \propto \tau^2$.) We also refer to the density fluctuation field $\delta(\vec{x}, \tau)$ and "peculiar" velocity field $\vec{v}(\vec{x}, \tau)$, which are related to the proper density $\rho$ and velocity $d\vec{r}/dt$ as follows:

$$\delta(\vec{x}, \tau) = \frac{\rho(\vec{x}, \tau)}{\bar{\rho}(\tau)} - 1 , \quad \vec{v}(\vec{x}, \tau) = \frac{d\vec{r}}{dt} - H\vec{r} = \frac{d\vec{x}}{d\tau} . \tag{2}$$

The reader will recognize $H$ as the Hubble parameter and should be familiar with its relation to the mean density $\bar{\rho}$ and the density parameter $\Omega = 8\pi G\bar{\rho}/(3H^2)$.

The density and velocity fields are assumed to evolve consistently with conservation of mass and momentum. For a nonrelativistic perfect fluid on scales much smaller than the Hubble distance $c/H$, this implies the fluid equations ([5]):

$$\text{Continuity:} \quad \frac{\partial \delta}{\partial \tau} + \vec{\nabla} \cdot [(1+\delta)\vec{v}] = 0 , \tag{3}$$

$$\text{Euler:} \quad \frac{\partial \vec{v}}{\partial \tau} + (\vec{v} \cdot \vec{\nabla})\vec{v} = -\frac{\dot{a}}{a}\vec{v} - \vec{\nabla}\phi \left[-\frac{1}{\rho}\vec{\nabla}p\right] , \tag{4}$$

$$\text{Poisson:} \quad \nabla^2 \phi = 4\pi G a^2 \delta\rho = \frac{3}{2}\Omega_0 H_0^2 a^{-1} \delta . \tag{5}$$

These are identical in form with the noncosmological fluid equations aside from the Hubble drag term in the Euler equation (N.B. $\dot{a}/a = aH$ is not the Hubble parameter!) and the fact that the mean density $\bar{\rho}$ does not contribute to the gravitational potential $\phi$. (In the literature $\phi$ is often called "peculiar," but this is carrying historical precedent too far. There is nothing peculiar about either $\phi$ or $\vec{v}$.) We assume that nongravitational (e.g., electromagnetic) forces are unimportant on large scales.

The fluid equations are valid for individual components (e.g., collisionless dark matter or baryons) as well as for the combined distribution of mass as long as the stress tensor is isotropic (i.e., given by the pressure $p$). Before trajectories intersect (or, it is believed, after the fluid variables are spatially averaged so that $\delta^2 \lesssim 1$), the fluid equations with $p = 0$ provide a good approximation to the dynamics of collisionless dark matter. For baryons, particle collisions are sufficiently rapid so that the stress is isotropic in the fluid frame and the pressure term must be included. Terms depending on pressure are kept in brackets in the following so that the reader can readily see how pressure modifies the evolution.

A key consequence of equation (4) is that gravity alone cannot generate vorticity $\vec{\omega} \equiv \vec{\nabla} \times \vec{v}$. Taking the curl of equation (4) we obtain

$$\frac{\partial \vec{\omega}}{\partial \tau} = \vec{\nabla} \times (\vec{v} \times \vec{\omega}) - \frac{\dot{a}}{a}\vec{\omega} \left[+\frac{1}{\rho^2}(\vec{\nabla}\rho) \times (\vec{\nabla}p)\right] . \tag{6}$$

The expression in brackets is called the baroclinic term. It vanishes for a pressureless or an isentropic fluid (i.e., one with constant specific entropy) or, more generally, for a barotropic fluid with $p = p(\rho)$.

The baroclinic term is very important in meteorology (it generates cyclones, tornados, and hurricanes!) but not cosmology. From equation (6) it follows that irrotational ($\vec{\omega} = 0$) flow remains irrotational in the absence of baroclinic torques (Kelvin's circulation theorem).

Equation (6) implies that any primeval vorticity decays as $\vec{\omega} \propto a^{-1}$ unless it is so large that turbulent stresses amplify vorticity through the $\vec{v} \times \vec{\omega}$ term. The latter possibility seems to be ruled out by the isotropy of the cosmic microwave background radiation. In this case, the velocity field prior to the epoch of galaxy formation (and today, on large scales) is expected to be a potential flow (i.e., irrotational). Potential flow is the essential assumption underlying the POTENT analysis ([3]). If $\vec{\omega} = 0$, we may write the velocity field in terms of a potential $\Phi$:

$$\vec{v} = -\vec{\nabla}\Phi \ , \quad \Phi(\vec{x}, \tau) = \Phi(0, \tau) - \int_0^{\vec{x}} \vec{v} \cdot d\vec{l} \ . \tag{7}$$

Any path may be used to evaluate the velocity potential $\Phi$, in particular, a radial path for which $\Phi(r, \theta, \phi) - \Phi(0) = -\int_0^r (cz' - H_0 r') dr'$. In the POTENT procedure ([3], [9], [4]), we first interpolate the redshifts and distance of galaxies in the local universe to define a smooth radial velocity field. The radial integral gives the velocity potential, whose transverse derivatives give the components of the smoothed velocity field not obtainable directly from redshifts and distances.

For a potential flow, the Euler equation may be reduced to an evolution equation for the velocity potential:

$$\text{Bernoulli:} \quad \frac{\partial \Phi}{\partial \tau} - \frac{1}{2}|\vec{\nabla}\Phi|^2 = -\frac{\dot{a}}{a}\Phi + \phi \left[+ \int \frac{dp}{\rho}\right] \ . \tag{8}$$

The Bernoulli, continuity, and Poisson equations are the basis for several approximate nonlinear solution methods ([25], [15], [20], [23], [1]).

The Eulerian fluid equations are commonly linearized to give the evolution at early times or on large scales such that $\delta^2 \ll 1$. Assuming a potential flow, the velocity field is described fully by its divergence, $\theta \equiv \vec{\nabla} \cdot \vec{v}$. Linearizing equation (3) and the divergence of (4), and assuming isentropic variations with $\vec{\nabla} p = (\partial p/\partial \rho)_S \vec{\nabla}\rho = c_s^2 \vec{\nabla}\rho$ (where $S$ is the specific entropy and $c_s$ is the sound speed), we get

$$\frac{\partial \delta}{\partial \tau} + \theta = 0 \ , \quad \frac{\partial \theta}{\partial \tau} = -\frac{\dot{a}}{a}\theta - 4\pi G\bar{\rho}a^2\delta - c_s^2 \nabla^2 \delta \ . \tag{9}$$

These two equations may be combined to give a second-order (in time) linear differential equation for $\delta$. The spatial dependence is solved using plane waves with comoving wavenumber $k$. It is easy to see that gravitational (Jeans) instability results if $k^2 c_s^2 < 4\pi G\bar{\rho}a^2$. For $c_s^2 = 0$ (or on any scales that are strongly Jeans unstable), the general solution for the density and velocity is a linear combination of growing and decaying modes $D_\pm(\tau)$:

$$\delta = \delta_+(\vec{x})D_+(\tau) + \delta_-(\vec{x})D_-(\tau) \ , \quad \theta = -\delta_+(\vec{x})\dot{D}_+(\tau) - \delta_-(\vec{x})\dot{D}_-(\tau) \ . \tag{10}$$

At late times the growing mode dominates. Its logarithmic derivative with respect to the expansion factor depends on the background cosmology chiefly through $\Omega$, and is written $d \ln D_+/d \ln a = f(\Omega) \approx \Omega^{0.6}$ ([27], [22]).

Sometimes it is stated that $\theta \propto \delta$ solely as a consequence of the continuity equation. This is false. If the linear growing mode dominates (and *if* the pressure, vorticity, and nongravitational forces are negligible), then $\theta = -aHf(\Omega)\delta$. However, if the decaying mode is present or any of the other conditions is violated, then $\theta$ is no longer proportional to $\delta$, despite mass conservation. Because the growing mode quickly overtakes the decaying mode, unless a perturbation is created with unnaturally small $\delta_+/\delta_-$, the growing mode should dominate by the present time so that independent measurements of $\delta$ and $\theta$ may be used to estimate $f(\Omega)$. (This is true even if the perturbations were created by nongravitational processes, provided that gravity subsequently drives $D_-/D_+ \to 0$.) Dekel *et al.* ([10]) have used a quasi-nonlinear generalization of this idea to place bounds on $\Omega$.

In the linear regime (i.e., while eqs. 9 are valid), the velocity and density of individual Fourier components evolve independently. When the perturbations become large different harmonics are coupled. The evolution is no longer local either in Fourier or real space. However, the evolution becomes somewhat easier to follow if we abandon the Eulerian description for a Lagrangian one.

## 3  Lagrangian description of trajectories

In a Lagrangian description one follows individual mass elements, in contrast with the Eulerian practice of tracking the values of the fluid variables at fixed spatial coordinates. Different mass elements (or particles) are labeled by a fixed Lagrangian coordinate $\vec{q}$, so that the trajectories are $\vec{x} = \vec{x}(\vec{q}, \tau)$. If there are no forces except gravity, the motion is governed by Newton's laws in comoving coordinates:

$$\frac{d^2\vec{x}}{d\tau^2} \equiv \left(\frac{\partial^2 \vec{x}}{\partial \tau^2}\right)_{\vec{q}} = -\frac{\dot{a}}{a}\frac{d\vec{x}}{d\tau} - \vec{\nabla}\phi \ . \tag{11}$$

This equation is to be solved for each mass element after we relate $-\vec{\nabla}\phi$ to the trajectories. How?

In general, $-\vec{\nabla}\phi$ must be computed by solving the Poisson equation with the mass distribution given by the instantaneous positions of all the mass elements. There are, however, several circumstances in which this computation simplifies.

First, if the mass distribution has sufficient symmetry, the Poisson equation may be solved using Gauss' theorem. For example, if the mass distribution is spherical about a point $\vec{x} = 0$, the solution is

$$-\vec{\nabla}\phi = -\frac{G}{ar^2}\left[M(r) - \bar{M}(r)\right]\vec{e}_r \ , \tag{12}$$

where $\bar{M}(r) = (4\pi/3)\bar{\rho}(ar)^3$, $r \equiv |\vec{x}|$, and $\vec{e}_r \equiv \vec{x}/r$. Before trajectories intersect, $M$ is fixed for a given mass element and is therefore a Lagrangian coordinate. Equations (11) and (12) together give an equation of motion for $r(\tau, M)$ for fixed $M$. The equation can be solved to give exact radial Keplerian trajectories. Identical results are obtained whether one uses Newton's laws in a noncosmological background or full general relativity ([27], §§19, 87).

Gauss' theorem may also be applied in the case of planar symmetry (and, of course, cylindrical symmetry). Rather than carrying out this derivation, we consider an alternative approach due to Zel'dovich, which is exact for plane-parallel perturbations of cold dust (collisionless matter with no thermal velocities) and provides a good approximation for small perturbations of arbitrary geometry.

First we choose $\vec{q}$ (without loss of generality) so that $\vec{x}(\vec{q}, 0) = \vec{q}$. Then, in general, $\vec{x}(\vec{q}, \tau) = \vec{q} + \vec{\psi}(\vec{q}, \tau)$, where $\vec{\psi}(\vec{q}, \tau)$ is called the *displacement field*. Mass conservation implies

$$\frac{1 + \delta(\vec{x}, \tau)}{1 + \delta(\vec{x}, 0)} = \sum_{\text{streams}} \left\|\frac{\partial \vec{x}}{\partial \vec{q}}\right\|^{-1} \ , \tag{13}$$

where the sum is taken over all the $\vec{q}$ present at a given $\vec{x}$. In the linear regime there is only one stream and the perturbations to the Jacobian determinant are small. Then, following Zel'dovich ([32]), we expand the Jacobian to first order in $\partial \psi^i / \partial x^j$. Assuming $\delta(\vec{x}, 0) = 0$, we get the following relation between $\delta$ and $\vec{\psi}$: $\delta = (4\pi G a^2 \bar{\rho})^{-1}\vec{\nabla}\cdot\vec{\nabla}\phi \approx -\vec{\nabla}_{\vec{q}}\cdot\vec{\psi}$, which yields (to first order in $\psi$)

$$\vec{\nabla}\phi \approx -4\pi G a^2 \bar{\rho}\,\vec{\psi} \ . \tag{14}$$

Substituting equation (14) into equation (11), one obtains a linear second-order ordinary differential equation in time for $\vec{\psi}(\vec{q}, \tau)$, whose general solution is

$$\vec{\psi}(\vec{q}, \tau) \approx D_+(\tau)\vec{\psi}_+(\vec{q}) + D_-(\tau)\vec{\psi}_-(\vec{q}) \ . \tag{15}$$

The same functions $D_\pm(\tau)$ appear here as in equation (10).

Zel'dovich proposed extending equation (15) into the nonlinear regime. Given the trajectories, it is straightforward to obtain the velocity and density fields as a function of the Lagrangian position $\vec{q}$, although more work is required to express them in terms of the Eulerian position $\vec{x} = \vec{q} - \vec{\psi}$. It is well known that the Zel'dovich approximation works very well until trajectories intersect and mass elements collapse to infinite density; it is superior to the linear Eulerian treatment in which mass elements never collapse. However, the Zel'dovich approximation breaks down after trajectories intersect. It is equivalent to a one timestep N-body algorithm in that particles are pushed in a fixed direction by an amount proportional to the initial acceleration, with no allowing for trajectories to reverse after crossing potential minima. There are various ways to cure this problem, including adding viscosity (converting the Euler equation to Burgers' equation), pre-filtering the density field, and using higher-order perturbation theory (equivalent to taking more than one timestep). We refer the interested reader to the review article of Shandarin & Zel'dovich ([29]) as well as to several contributions in this volume (and [7], [16], [21], [14]) for further discussion of the Zel'dovich approximation and extensions.

## 4  Lagrangian fluid equations

The Lagrangian approach may be applied to give equations of motion for fluid variables in addition to trajectories. We will derive the Lagrangian fluid equations for a pressureless gas beginning from the Eulerian equations. The same results follow for cold dust by considering trajectories of adjacent freely-falling mass elements.

The procedure we follow is to rewrite the fluid equations using the Lagrangian (or convective) time derivative

$$\frac{d}{d\tau} \equiv \frac{\partial}{\partial\tau} + \vec{v} \cdot \vec{\nabla} \ . \tag{16}$$

Applied to any Eulerian field $f(\vec{x}, \tau)$, $df/d\tau$ gives the time derivative following the fluid since the fluid velocity is $\vec{v} = d\vec{x}/d\tau$.

Replacing the Eulerian time derivatives by Lagrangian ones, the zero-pressure fluid equations in comoving coordinates become ([6], but note that the present definition of $\vec{\omega}$ differs by a factor of 2)

$$\text{Continuity:} \quad \frac{d\delta}{d\tau} + (1+\delta)\theta = 0 \ , \tag{17}$$

$$\text{Raychaudhuri:} \quad \frac{d\theta}{d\tau} + \frac{\dot{a}}{a}\theta + \frac{1}{3}\theta^2 + \sigma^{ij}\sigma_{ij} - \frac{1}{2}\omega^2 = -4\pi G a^2 \bar{\rho}\delta \ , \tag{18}$$

$$\text{Vorticity:} \quad \frac{d\omega^i}{d\tau} + \frac{\dot{a}}{a}\omega^i + \frac{2}{3}\theta\omega^i - \sigma^i{}_j\omega^j = 0 \ , \tag{19}$$

$$\text{Shear:} \quad \frac{d\sigma_{ij}}{d\tau} + \frac{\dot{a}}{a}\sigma_{ij} + \frac{2}{3}\theta\sigma_{ij} + \sigma_{ik}\sigma^k{}_j + \frac{1}{4}\omega_i\omega_j - \frac{1}{3}\delta_{ij}\left(\sigma^{kl}\sigma_{kl} + \frac{\omega^2}{4}\right) = -E_{ij} \ , \tag{20}$$

where we have defined the *shear* and *tide* tensors:

$$\sigma_{ij} \equiv \frac{1}{2}\left(\frac{\partial v_i}{\partial x^j} + \frac{\partial v_j}{\partial x^i}\right) - \frac{1}{3}\theta\delta_{ij} \ , \quad E_{ij} \equiv \frac{\partial^2\phi}{\partial x^i \partial x^j} - \frac{1}{3}(\nabla^2\phi)\delta_{ij} \ . \tag{21}$$

(Note that repeated indices are to be summed over and that we are implicitly assuming the use of Cartesian comoving coordinates.) The shear and tide tensors are symmetric and traceless.

It is remarkable that, aside from $E_{ij}$, equations (17)–(20) provide a closed set of local evolution equations for $\delta$, $\theta$, $\vec{\omega}$, and $\sigma_{ij}$, with no spatial derivatives aside from those implicit in the tide tensor. Thus, in the absence of gravity, each mass element evolves independently of all the others, at least until it crosses other elements (recall that we have neglected pressure). Note that in the linear regime, equations (17) and (18) reduce to (9) (neglecting pressure) because $d/d\tau \approx \partial/\partial\tau$. However, the Lagrangian equations have an advantage in retaining a local form in the nonlinear regime.

The locality of the fluid equations inspires us to go further to try to obtain an evolution equation for $E_{ij}$. In the Newtonian framework this is unnatural (though not illegal!): gravity is given instantaneously by solution of the Poisson equation and not by some time evolution equation. However, in general relativity it is natural to treat the tide tensor on the same footing as the fluid variables. Ellis ([12], [13]) derives the appropriate equations using the Bianchi identities and Einstein equations. Written in comoving coordinates, the evolution equation for $E_{ij}$ is

$$\frac{dE_{ij}}{d\tau} + \frac{\dot{a}}{a}E_{ij} + \theta E_{ij} - 3\sigma^k{}_{(i}E_{j)k} + \delta_{ij}\sigma^{kl}E_{kl} + \frac{1}{2}\epsilon^{kl}{}_{(i}E_{j)k}\omega_l - \nabla_k\epsilon^{kl}{}_{(i}H_{j)l} = -4\pi G a^2 \bar{\rho}(1+\delta)\sigma_{ij} \ . \tag{22}$$

Parentheses indicate symmetrization: $\sigma^k{}_{(i} E_{j)k} \equiv \frac{1}{2} \left( \sigma^k{}_i E_{jk} + \sigma^k{}_j E_{ik} \right)$. The fully antisymmetric Levi-Civita tensor is $\epsilon_{ijk}$ (with $\epsilon_{123} = +1$). Equation (22) can be derived (with some difficulty) as an exact equation in the Newtonian limit ([17]). The difficult part is the term involving the symmetric traceless tensor $H_{ij}$, called, in general relativity, the *magnetic part of the Weyl tensor*. Ellis says that $H_{ij}$ has no Newtonian analogue, but that is not completely correct, although its Newtonian interpretation at present remains unclear. On the other hand, the tide tensor $E_{ij}$, known in general relativity as the electric part of the Weyl tensor, is easily understood in the Newtonian framework. The Weyl tensor is the traceless part of the Riemann tensor (the tensor responsible for Newtonian tidal forces). The electromagnetic terminology is used because $E_{ij}$ and $H_{ij}$ obey equations similar to the Maxwell equations. For example, equation (22) is analogous to the Ampère law $\partial \vec{E}/\partial t + \vec{\nabla} \times \vec{B} = 4\pi \vec{J}$.

Aside from the term involving $H_{ij}$, equation (22) is purely local. Thus, if $H_{ij} = 0$, we have obtained a closed set of local Lagrangian equations for the nonlinear evolution of cold dust. This fact was first noted by Barnes & Rowlingson ([2]) and applied in cosmology by Matarrese, Pantano, & Saez ([24]). Bertschinger & Jain ([6]) fully explored the consequences of this assumption for nonlinear evolution of cold dust in an Einstein-de Sitter universe. All of these authors assumed that if $\vec{\omega} = 0$ and $\sigma_{ij} \propto E_{ij}$ initially, then $H_{ij} = 0$ at least until trajectories intersect. The physical interpretation is that local Newtonian evolution is causal and exact in the weak-field limit if mass motions do not generate gravitomagnetism or gravitational radiation.

However, recent analytical and numerical results suggest that $H_{ij}$ in equation (22) does not vanish identically in the Newtonian limit. The algebraic complexity of the equations makes it difficult to analyze the behavior of the Weyl tensor in the Newtonian limit. Nevertheless, neglecting $H_{ij}$ may provide a good approximation in many circumstances, and we will follow the consequences of this assumption in Section 7 below.

## 5 Four propositions

Consider the evolution of irrotational, pressureless matter under gravity. Would you agree with the following four propositions?

<u>Proposition A</u>: For a given $\delta$ and $\dot{\delta}$, a spherical tophat perturbation is the configuration that collapses most rapidly.

<u>Proposition B</u>: Initial density maxima are the sites where collapse first occurs.

<u>Proposition C</u>: Underdense regions do not collapse before colliding with other streams.

<u>Proposition D</u>: The final stage of collapse is generically one-dimensional, leading to Zel'dovich pancakes.

Now consider four alternative propositions.

<u>Proposition 1</u>: For a given $\delta$ and $\dot{\delta}$, a spherical tophat perturbation is the configuration that collapses most *slowly*.

<u>Proposition 2</u>: Initial density maxima are *not* the sites where collapse first occurs.

<u>Proposition 3</u>: Underdense regions *do* collapse before colliding with other streams, *if* the initial shear is not too small.

<u>Proposition 4</u>: The final stage of collapse is generically *two*-dimensional, leading to strongly prolate filaments.

Propositions 1–3 are true while A–C are false. Moreover, if the magnetic part of the Weyl tensor is negligible, Proposition 4 is true also. In the following we will see show how Propositions 1–4 follow from the Lagrangian fluid equations.

# 6 Collapse theorem and corollaries

Combining the Lagrangian continuity and Raychaudhuri equations we obtain the following exact equation for cold dust:

$$\ddot{\delta} + \frac{\dot{a}}{a}\dot{\delta} = \frac{4}{3}\frac{\dot{\delta}^2}{1+\delta} + (1+\delta)\left(\sigma^{ij}\sigma_{ij} - \frac{1}{2}\omega^2 + 4\pi Ga^2\bar{\rho}\,\delta\right) \ . \tag{23}$$

This equation makes no assumptions about $\vec{\omega}$ or $H_{ij}$; it is independent of the Weyl tensor. We can use it to investigate gravitational collapse ($\delta \to \infty$) of mass elements before their trajectories intersect others.

If $\vec{\omega} = 0$, equation (23) shows that collapse is accelerated by nonzero shear ($\sigma^{ij}\sigma_{ij}$ is non-negative). The spherical tophat perturbation, with $\delta(\vec{x}, \tau_i) = \delta_i$ for $r < R$ and 0 for $r > R$, has uniform radial flow for $r < R$, with $\vec{v} \propto \vec{x}$, so that $\sigma_{ij} = 0$. Therefore, for a given $\delta$ and $\dot{\delta}$, the spherical tophat perturbation collapses more slowly than any other configuration (Proposition 1, now a Theorem). The physical explanation is that shear increases the rate of growth of the convergence of fluid streamlines.

<u>Corollary 1</u>: If $\delta > 0$, an irrotational growing-mode perturbation collapses in finite time ([6]).

<u>Corollary 2</u>: The collapse time depends on all three initial eigenvalues of $\partial^2\phi/\partial x^i\partial x^j$ (equivalently, for growing mode perturbations, $\partial^2\Phi/\partial x^i\partial x^j$), not just on the trace (i.e., $\delta$). Therefore, local maxima of $\delta$ are not, in general, local minima of the collapse time (Proposition 2).

<u>Corollary 3</u>: If $\sigma_{ij} \neq 0$, a perturbation may collapse even if $\delta \leq 0$ initially (Proposition 3). This follows from a continuity argument: If $\delta = 0$ initially, a zero-shear perturbation barely avoids collapse in finite time. Nonzero shear induces a positive $\dot{\delta}$, speeding up collapse, which then occurs in a finite time.

Propositions 1–3 also follow under the Zel'dovich approximation, where the density depends on all three eigenvalues of the strain tensor $\partial x^i/\partial q^j$. However, the Zel'dovich approximation used to evaluate this tensor is only an approximation, whereas equation (23) is exact.

# 7 Results of local evolution

Proposition 4 is difficult to analyze because the geometry of collapse depends on the shear tensor, whose evolution depends on the tide. In the Zel'dovich approximation, collapse is purely kinematical: because the displacements are proportional to the initial accelerations, collapse occurs first in the direction corresponding to the smallest eigenvalue of the initial strain tensor. (Not all points will collapse because in some places all three eigenvalues of $\partial\psi^i/\partial q^j$ will be positive.) However, this is not exact because the Zel'dovich approximation neglects the gravitational feedback of the collapsing flow on itself.

Bertschinger & Jain ([6], see also [8]) have analyzed equation (22) and shown that the $-3\sigma^k{}_{(i}E_{j)k}$ shear-tide coupling term drives the evolution toward prolate configurations of the shear and tides. If $H_{ij} = 0$, the shear tensor generically becomes strongly prolate (with two negative and one positive eigenvalue) as collapse is approached. Note that this does not imply that the instantaneous shape of a collapsing object is prolate, because $\partial x^i/\partial q^j$ is related to the time integral of the velocity gradient tensor. Nevertheless, the result is surprising because it contradicts the Zel'dovich pancake paradigm.

There is a plausible physical explanation for prolate collapse. The gravitational binding energy is larger for a linear mass distribution (with $\phi \propto \ln r \to -\infty$ for a line mass) than for a planar mass distribution ($\phi \to$ constant as $r \to 0$ for a sheet of mass). Although the gravitational energy is still larger for a one-dimensional (spherical) collapse, nonzero shear and angular momentum conservation prevent spherical collapse in general.

Assuming $H_{ij} = 0$, Bertschinger & Jain also showed that for a growing-mode Gaussian random initial density field in an Einstein-de Sitter universe, 56% of initially underdense cold dust mass elements collapse. Only 22% of the mass can avoid collapse before intersecting other mass elements.

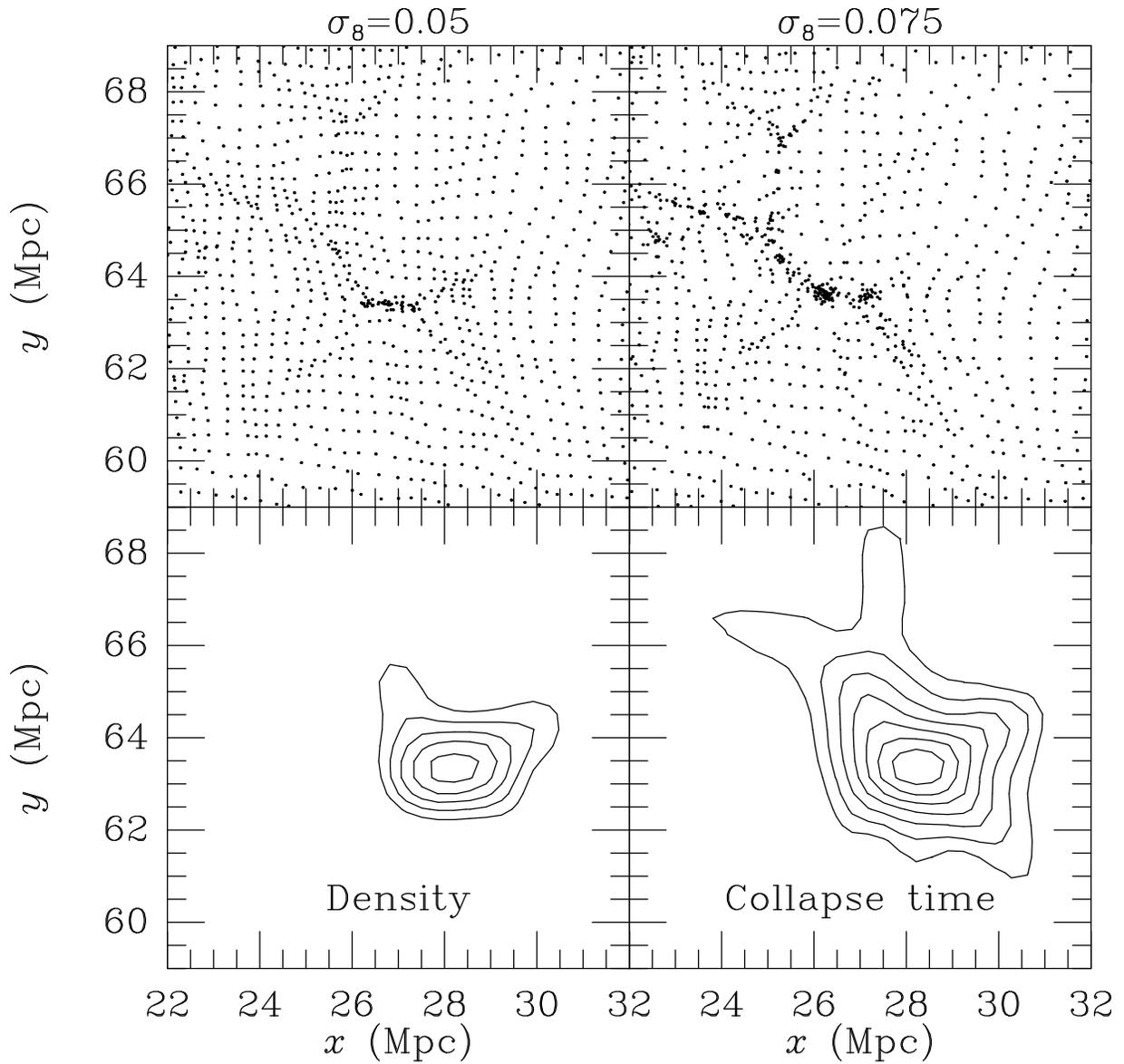

Figure 1: Top: particles in one Lagrangian slice of a high-resolution cold dark matter N-body simulation at times specified by the specified linear amplitude $\sigma_8$. The Eulerian $x$ and $y$ comoving positions are shown for particles in a single Lagranian layer $q_z = $ constant. Note how filamentary the structures are; the extent in the third dimension is small. Bottom: smoothed contours of constant initial $\delta$ and collapse time, the latter computed assuming that the magnetic part of the Weyl tensor vanishes.

Recently, S. White has investigated numerically the collapse of a uniform ellipsoidal overdense perturbation embedded in an Einstein-de Sitter universe. His numerical results agree well with approximate analytical theory neglecting the tidal feedback with the surroundings ([31]) and disagree quantitatively with the predictions of the Bertschinger-Jain theory assuming $H_{ij} = 0$. Assuming there is no subtle error in the simulation, one is forced to conclude $H_{ij} \neq 0$. This is difficult to check analytically, and the status of $H_{ij}$ remains unclear, although it is known that the $H_{ij}$ term in equation (22) vanishes in cases of high symmetry. However, in the general case it seems appropriate to regard the neglect of $H_{ij}$ as an approximation whose range of validity requires further investigation.

The $H_{ij} = 0$ predictions are tested in Figure 1 against a high-resolution (20 kpc comoving force softening distance) N-body simulation of the standard cold dark matter model with $288^3$ particles in a 100 Mpc cube. The first structure to collapse in the simulation was found at high redshift ($\sigma_8 = 0.05$ corresponds to $z = 20$ if the model is normalized to the COBE quadrupole). The structure in Figure 1 lies nearly in the $x$-$y$ plane, indicating that the collapse is strongly filamentary, as predicted. At a later time ($\sigma_8 = 0.075$) the strongest filament breaks into two clumps, which subsequently merge. Contours of constant linear $\delta$ are shown in the bottom left panel; if collapse occurs when $\delta = 1.686$ as predicted by the spherical tophat model, then the $\delta$ contours correspond to collapses occurring at $1 + z_c = (12, 14, 16, 18, 20)$. (These are underestimates because the density field was smoothed slightly to produce the contours.) The lower right panel shows the results for $1 + z_c$ (with contour levels $12, 14, 16, 18, 20, 22, 24$) computed by solving equations (17)–(22) assuming $H_{ij} = 0$. (The $\sigma_8$ labels apply only to the top two panels and not the bottom ones.) Note that both sets of contours are plotted in Lagrangian space; they purport to show (under the approximations of the spherical tophat model or $H_{ij} = 0$) the Lagrangian volumes that should have collapsed by a given redshift. The point of initial collapse is rather close to the density maximum in Lagrangian space (in this case Proposition B is nearly correct), but both are shifted relative to the point of Eulerian collapse by long wavelength displacements. One can see that the spherical tophat model predicts a later collapse (supporting Proposition 1, the Collapse Theorem), with less material having collapsed by a given redshift (because of its neglect shear), and it does not suggest the formation of filaments.

This test is not very precise, but it shows that the local description makes some predictions that are close to what happens in this N-body simulation. Given the high resolution of the simulation, the strong small-scale power of the cold dark matter spectrum, and the strong nonlinearity of the density distributions, Figure 1 must be counted a real success of the Lagrangian fluid description.

## 8 Conclusions

Gravitational instability theory still has surprises waiting to be uncovered. Eulerian theory is well developed but rather stale, while studies based on Lagrangian trajectories remain fertile. The recent applications of Lagrangian fluid dynamics (not SPH!) to cosmology have begun to attack fully nonlinear problems that previously could be addressed only in cases of high symmetry. The completeness of this approach remains unclear, however, as long as the Newtonian (and relativistic!) behavior of $H_{ij}$ (the magnetic part of the Weyl tensor) is not understood. It seems remarkable that basic questions about Newton's laws still exist, but this fact only serves to demonstrate the richness of nonlinear gravitational clustering.

Lagrangian methods are not a panacea. Even if the local description of the fluid variables prior to trajectory-crossing were to prove correct, the description is incomplete without knowledge of the Eulerian location of the mass elements. This fact suggests that it may prove fruitful to combine Lagrangian fluid dynamics with integration of the trajectories to supplement, or even replace ([24]), standard N-body techniques.

**Acknowledgements.** I have benefited from the insightful comments of Bhuvnesh Jain, Andrew Hamilton, and Simon White. This work was supported by NSF grant AST90-01762.